\documentclass[prl,twocolumn,superscriptaddress]{revtex4}
\usepackage[dvips]{graphicx}
\usepackage{latexsym}
\usepackage{epsfig}
\usepackage{bm}

\begin{document}

\renewcommand{\ni}{{\noindent}}
\newcommand{\dprime}{{\prime\prime}}
\newcommand{\be}{\begin{equation}}
\newcommand{\ee}{\end{equation}}
\newcommand{\bea}{\begin{eqnarray}} 
\newcommand{\eea}{\end{eqnarray}}
\newcommand{\nn}{\nonumber} 
\newcommand{\bk}{{\bf k}}
\newcommand{\bQ}{{\bf Q}}
\newcommand{\q}{{\bf q}}
\newcommand{\s}{{\bf s}}
\newcommand{\bN}{{\bf \nabla}}
\newcommand{\bA}{{\bf A}}
\newcommand{\bE}{{\bf E}}
\newcommand{\bj}{{\bf j}}
\newcommand{\bJ}{{\bf J}}
\newcommand{\bs}{{\bf v}_s}
\newcommand{\bn}{{\bf v}_n}
\newcommand{\bv}{{\bf v}} 
\newcommand{\la}{\langle}
\newcommand{\ra}{\rangle} 
\newcommand{\dg}{\dagger}
\newcommand{\br}{{\bf{r}}} 
\newcommand{\brp}{{\bf{r}^\prime}} 
\newcommand{\bq}{{\bf{q}}}
\newcommand{\hx}{\hat{\bf x}} 
\newcommand{\hy}{\hat{\bf y}}
\newcommand{\bS}{{\bf S}} 
\newcommand{\cU}{{\cal U}}
\newcommand{\cD}{{\cal D}} 
\newcommand{\bR}{{\bf R}}
\newcommand{\pll}{\parallel} 
\newcommand{\sumr}{\sum_{\vr}} 
\newcommand{\cP}{{\cal P}} 
\newcommand{\cQ}{{\cal Q}} 
\newcommand{\cS}{{\cal S}}
\newcommand{\ua}{\uparrow} 
\newcommand{\da}{\downarrow}

\title{Inhomogeneous metallic phase upon disordering a two dimensional Mott
insulator}
\author{Dariush Heidarian}
\affiliation{Department of Theoretical Physics, Tata Institute of 
Fundamental Research, Mumbai 400 005, India}
\author{Nandini Trivedi}
\affiliation{Department of Theoretical Physics, Tata Institute of 
Fundamental Research, Mumbai 400 005, India}
\affiliation{Department of Physics, University of Illinois at 
Urbana-Champaign, IL 61801}
\begin{abstract}
\vspace{0.1cm}
We find that isoelectronic disorder
destroys the spectral gap
in a Mott-Hubbard insulator in 2D leading, most unexpectedly,
to a new metallic phase.
This phase is spatially inhomogeneous with metallic behavior coexisting with
antiferromagnetic long range order. Even though
the Mott gap in the pure system is much larger than antiferromagnetic exchange,
the spectral gap is destroyed locally in regions where the
disorder potential is high enough to overcome the inter-electron
repulsion thereby generating
puddles where charge fluctuations are enhanced. With increasing disorder,
these puddles expand and concomitantly
the states at the Fermi energy get extended
leading to a metallic phase.
We discuss the implications of our results for experiments.
\typeout{polish abstract}
\end{abstract}

\maketitle

Disorder can lead to new important phenomena:
such as localization in disordered media\cite{gangof4,lee-tvr}, 
metal-insulator transitions\cite{mit_review},
precisely quantized Hall plateaus in the quantum Hall effect, 
and zero resistance from pinned vortices in a type II superconductor.
In systems as diverse as 
superconducting films\cite{ghosal}, high $T_c$
cuprates\cite{davis} and manganites\cite{dagotto} disorder 
produces nanoscale inhomogeneities which can have a profound influence 
on the properties of these systems.

We propose the existence of a new metallic phase at $T=0$
which is generated when potential disorder is introduced into a Mott
insulator. This metallic phase is highly inhomogeneous and coexists with
long range antiferromagnetic order. It is sandwiched between two qualitatively
different insulators: at low disorder,
a Mott insulator with a spectral gap and
antiferromagnetic (AFM) long range order,
and, at high disorder,
a paramagnetic Anderson insulator with gapless excitations.
Our result is surprising since one would expect that the
primary role of disorder is to
localize wave functions, which is opposite to the conditions for
forming extended states necessary for metallic behavior.

We find that the route to the formation of a metal at intermediate disorder
is the following:
Given a random potential profile, there are two types of regions:
weakly disordered (WD) with relatively small potential fluctuations
and strongly disordered (SD) with larger fluctuations.
AFM order exists in the WD regions that consist of 
singly occupied sites with a large local Mott gap.
On the other hand, in the SD regions the local density deviates from
one per site which disrupts the antiferromagnetic ordering
and also leads to the collapse of the gap locally.
There is thus a `phase separation' between the
insulating AFM WD regions and the metallic SD regions with suppressed AFM.
The low lying 
excitations of the system remain trapped in the SD regions.
With increasing disorder the SD regions grow and 
along with that, the eigenfunctions
get extended throughout the system and the system becomes metallic
at intermediate disorder. At high disorder the system 
becomes a paramagnetic Anderson insulator.
We substantiate the above picture with calculations of the magnetic 
correlations, the local density of states and the optical conductivity.

We propose that experiments on cuprates in their parent Mott insulating
phase, could introduce disorder by isoelectronic substitution, or by
bombardment, without changing
the carrier doping. Such experiments
should be able to see this unusual metallic phase at intermediate disorder.
Recently, Vajk et. al\cite{vajk}
have measured the spin correlations using neutron scattering 
in $La_2CuO_4$ with
$Zn$ substitution on the $Cu$ sites. 
It will be necessary to complement the magnetic information from neutron
scattering with transport and spectroscopy to look for signatures
of an inhomogeneous metallic phase.

\noindent {\it Model and Calculation:}
We model the 2D disordered Mott insulator by a
repulsive Hubbard model at half filling (one electron per site)
with potential disorder:

\begin{equation}
{\cal H} = {\cal K} + U\sum_{i} n_{i \uparrow} n_{i \downarrow}
 + \sum_{i,\sigma} (V_{i}-\mu ) n_{i\sigma}.
\label {hamil}
\end{equation}
${\cal K} = -t\sum_{<ij>,\sigma} (c_{i\sigma}^{\dag} c_{j\sigma} + h.c.)$
is the kinetic energy,
$c_{i\sigma}^{\dag}$ ($c_{i\sigma}$) the creation (destruction)
operator for an electron with spin $\sigma$ on a site ${\bf r}_i$ of
a square lattice of $N$ sites,
$t$ the near-neighbor hopping,
$U$ is the on-site repulsion between electrons,
$n_{i\sigma}= c_{i\sigma}^{\dag}c_{i\sigma}$, and
$\mu$ the chemical potential.
The random potential $V_{i}$ is chosen independently
at each ${\bf r}_i$ and is uniformly distributed in the interval
$[-V,V]$;
$V$ thus controls the strength of the disorder.
All energies are measured in units of $t$.
This is a minimal model containing
the interplay of electronic correlations and localization:
for zero disorder $V=0$ it describes a Mott insulator at half filling,
and for $U=0$ it reduces to the (non-interacting)
Anderson localization problem.
We report results for $U=4t$ and lattices up to $50\times 50$.

The dynamical mean field theory\cite{dmft} has been widely
used to study Mott insulators 
and includes the quantum fluctuations associated 
with an effectively single site approximation of the system.
The inhomogeneous Hartree Fock approximation that we implement here is a 
complementary method which does not contain the dynamical fluctuations but,
when solved self consistently as we do here, 
captures the inhomogeneity in the spatial structure of the electronic 
density, a feature which is at the heart of problems with disorder.
The only known way to include both the spatial and temporal 
fluctuations is quantum Monte Carlo (QMC) but that method too has limitations 
arising from finite size effects, inability to extract dynamical information 
without the use of complicated maximum entropy methods and the sign problem 
for  fermions that limits the calculations to finite temperatures. 
\begin{figure}
\begin{center}
\vskip-2mm
\hspace*{1mm}
\epsfig{file=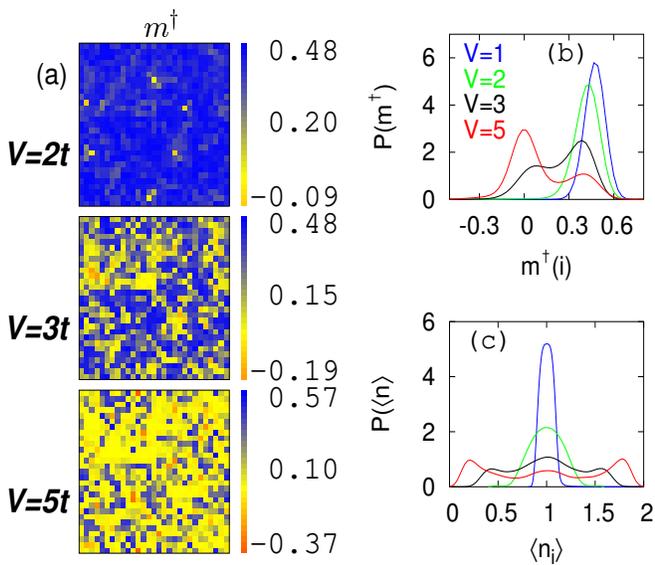,height=3.0in,width=3.5in,angle=0}
\vskip1mm
\caption{
(a) Left panels show the local staggered magnetization 
$m^\dagger(i)\equiv\langle S_z (i) \rangle (-1)^{i_x+i_y}$. The regions in blue have AFM order; the defective sites with reduced AFM order
are shown in yellow and red. 
The data is for the disordered Hubbard model at half filling 
with $U=4t$ on a $28\times 28$ lattice for disorder strengths 
$V=2t$; $V=3t$; $V=5t$
for one realization of disorder. 
(b) Probability distribution $P(m^\dagger)$ of $m^\dagger(i)$
for different values of $V$. For $V=1$, $P(m^\dagger)$ has a peak near 0.4.
With increasing
$V$, $P(m^\dagger)$ gets broader and develops weight near 0 indicating the 
growth of paramagnetic regions.
(c) Probability distribution $P(n)$ of the site-occupancy 
$\langle n_i\rangle$
for different values of $V$. For $V=1$, $P(n)$ has a peak near an
average single site occupancy $\langle n_i\rangle\approx 1 $. With increasing
$V$, P(n) gets broader and develops weight for doubly occupied and
unoccupied sites. 
}
\label{fig1}
\end{center}
\end{figure}
We begin by treating the spatial fluctuations of the 
local densities 
$ n_{i\sigma}=\langle c_{i\sigma}^\dagger c_{i\sigma}\rangle$ and
local magnetic fields
$ h_i^+=-U\langle c_{i\ua}^\dagger c_{i\da}\rangle$
and $ h_i^-=-U\langle c_{i\da}^\dagger c_{i\ua}\rangle$
using a {\it site dependent} mean field approximation.
As a consequence,
we get an effective Hamiltonian which is quadratic in the fermion
operators and can be diagonalized for a system with periodic 
boundary conditions. 
Starting with an initial guess for $h^\pm_i$'s and 
$n_{i\sigma}$'s\cite{footnote_mu} 
we numerically solve for the eigenvalues $\epsilon_n$ and eigenvectors
$\psi_{n\sigma}({\bf r}_i)$
of a $2N\times 2N$ matrix.
The local fields are then determined in terms of the eigenfunctions
and the eigenvalue problem with these new local fields as input is then 
iterated until {\it self consistency is achieved at each site}.

%
\begin{figure}
\begin{center}
\vskip-2mm
\hspace*{1mm}
\epsfig{file=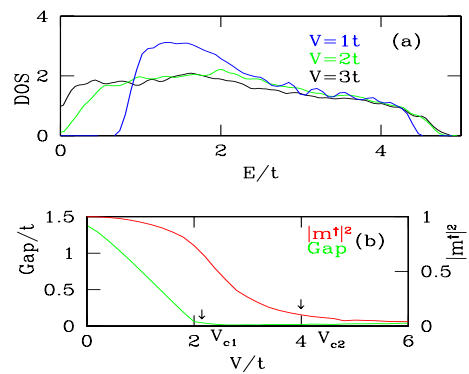,height=2.0in,width=2.5in}
\vskip1mm
\caption{
(a) Density of states averaged over 10 realizations at half filling and $U=4t$
for $V/t=1,2,3$. 
For $V=t$ there is a Mott gap in the spectrum at the 
chemical potential $E=0$;
and for $V=2t$ the gap has closed.
(b) Single particle energy gap as a function of $V$ showing
the collapse of the gap at $V_{c1}\sim 2.2t$ (scale on LHS).
The decrease of the AFM staggered order parameter $m^\dagger$ (scale on RHS),
as a function 
of disorder strength $V$ and its vanishing beyond $V_{c2}\sim 4t > V_{c1}$. 
Note that at $V=2t$ the system is still 
strongly AFM even though the Mott gap has vanished.
}
\label{fig2}
\end{center}
\end{figure}
\noindent{\it Antiferromagnetic Order and Spectral Gap:}
Fig.~1 shows that the local staggered magnetization 
$m^\dagger(i)\equiv (-1)^{i_x+i_y}\la S_z(i) \ra$ 
is largely antiferromagnetic for $V=2t$ 
with a few local defective regions. 
With increasing disorder the defective regions having 
reduced antiferromagnetic order grow in size. 

We next look at the behavior of the two defining characteristics of a 
Mott insulator in Fig.~2: the spectral gap obtained from the lowest eigenvalue 
of the effective Hamiltonian, and the AFM order parameter $m^\dagger(i)$ 
obtained from the spatial behavior of the 
spin-spin correlation function at large distances,
as a function of disorder. The first surprise is that 
even though the energy scale for charge fluctuations is $U\gg J\sim t^2/U$, 
the scale for antiferromagnetic exchange, the spectral gap vanishes at 
$V_{c1}\approx 2.2t$ which is 
lower than the critical disorder $V_{c2}\approx 4t$ where AFM long 
range order (AFLRO) vanishes.

The occurrence of two critical disorder strengths defines three
distinct regions:
In region I defined by $0\le V \le V_{c1}$ the system is
a Mott insulator with a finite gap and AFLRO.
The intermediate region II, $V_{c1}< V < V_{c2}$ 
is extremely unusual with AFLRO but no gap and is 
discussed in greater detail below.
In region III, $V \ge V_{c2}$
the system is simply an Anderson or localized insulator with gapless 
excitations and no magnetic order.

\noindent{\it Nature of eigenfunctions:}
We argue below
that in region II the system is a disordered AFM metal.
How does this metal come about?
The spatial extent of
the eigenstates near the
Fermi energy in Fig.~3(a) distinctly shows that 
they are
localized at low $V=2t$ and high disorder $V=5t$, but quite surprisingly at 
intermediate disorder $V=3t$ the states get more extended.
\begin{figure}
\begin{center}
\vskip-2mm
\hspace*{1mm}
\epsfig{file=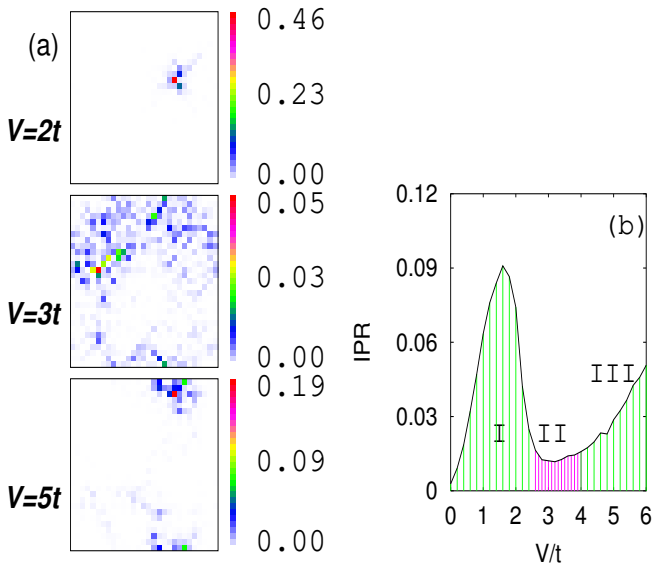,height=3.0in,width=3.5in,angle=0}
\vskip1mm
\caption{
(a) Single particle eigenstates at the Fermi energy for 
disorder strengths $V=2t,3t,5t$.
Note that the eigenstates are localized for low $V=2t$ and high disorder $V=5t$
but surprisingly are more extended at intermediate disorder of $V=3t$.
(b) Inverse participation ratio (IPR) or the localization length 
$\xi_{loc}^{-2}$ for the wave function at the Fermi energy
as a function of disorder $V$. In both regions I and III 
IPR increases with disorder or $\xi_{loc}$ decreases with disorder, as 
expected. But in the intermediate region II, the wave function in fact gets 
less localized with increasing disorder.
}
\label{fig3}
\end{center}
\end{figure}
The localization length 
$\xi_{{\rm loc}}(\alpha)$
for the (normalized) state $\psi_{\alpha}({\bf r}_i)$
is related to the inverse participation ratio 
$IPR\equiv \sum_{{\bf r}_i}|\psi_{\alpha}({\bf r}_i)|^4 
\propto \xi^{-2}_{{\rm loc}}(\alpha)$~\cite{lee-tvr}.
One would expect that with increasing disorder, $\xi_{\rm loc}$
would decrease or correspondingly IPR would increase. Instead we see 
in Fig.~3(b) 
a very definite decrease in IPR for $V\approx U/2$ signaling that even though
disorder is increasing the states are getting more extended. 
Such anomalous behavior continues till about $V\approx 3U/4$ and then 
once again reverts to the 
usual behavior where IPR increases with $V$.
Based on numerical calculations on finite size systems it is 
difficult to prove if the states in region II are truly extended, but 
what is abundantly clear is that the IPR does show a very definite 
non-monotonic behavior signaling a marked change in the nature of 
the states, at least on mesoscopic scales.\cite{footnote_states}.

Previous QMC simulations 
on disordered bosonic\cite{bhm} and fermionic\cite{fhm} 
have shown that potential disorder that breaks particle-hole symmetry
produces superfluid stiffness or conductivity 
which can increase with disorder. At a Hartree Fock level such an effect 
can be understood as a screening of the random potential by the interactions 
between the particles\cite{screening}. 
Our calculations go beyond simple screening arguments and 
have revealed the physical origin of the anomalous behavior seen in Fig.~3
at a microscopic level. 

\noindent{\it Frequency-dependent conductivity:}
The nature of the frequency dependent conductivity 
$Re \sigma(\omega)={\rm Im} \Lambda(\omega)/\omega$
gives insight into the conducting properties of the phases.
$\Lambda(\omega)$ is the 
Fourier transform of
$\Lambda(\tau)= \langle j(\tau)j(0) \rangle$, the
(disorder averaged) current-current correlation function.
As shown in Fig.~4 
the low frequency behavior of 
$\omega\sigma(\omega)$ has a finite gap
in the Mott region I for $V=t$; shows a linear $\omega$ dependence in the 
metallic region II for $V=3t$ which implies that there 
is a finite dc conductivity $\sigma(\omega\rightarrow 0)$;
and shows a $\omega^3$ dependence 
in the Anderson insulating regime III for $V=5t$,
which implies that there are gapless excitations but nevertheless
$\sigma\rightarrow 0$. 
\begin{figure}
\begin{center}
\vskip-2mm
\hspace*{1mm}
\epsfig{file=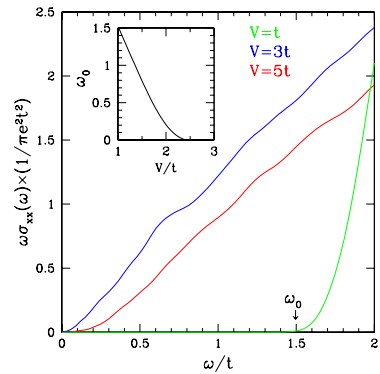,height=2.0in,angle=0}
\vskip1mm
\caption{
Frequency dependence of the conductivity 
for $V=1t$ in region I 
showing a gap $\omega_0$ 
in the joint density of states;
for $V=3t$ in region II showing a linear dependence 
indicative of metallic behavior; and for $V=5t$ in region III showing
$\omega^3$ dependence indicative of Anderson localization.
The inset shows the decrease of the gap $\omega_0$
with increasing 
disorder $V$ and its vanishing around $V_{c1}=2.2t$.
}
\label{fig4}
\end{center}
\end{figure}

\noindent{\it Origin of the Metallic Phase:}
Now that we have seen evidence for 
a metallic phase at 
intermediate disorder, the question arises as to how such a phase comes 
about. 

In the clean system, the repulsive interactions prevent double occupancy. 
But once there is potential disorder, electrons can hop into sites that 
have deep potential wells and thereby lower their energy.
This leads to
a broadened 
distribution of local occupancy 
and a growth
of paramagnetic sites, as illustrated in Fig.~1.
In the classical limit, 
sites with a large positive
potential $V_i>U/2$ are unoccupied; 
sites with $-U/2<V_i<U/2$ are singly occupied; and 
sites with large negative $V_i<-U/2$ are doubly occupied. Only sites
with one electron can contribute to AFM order if they form a connected
cluster. 
The critical probability $p_c$ for classical percolation
of vacancies on a 2D square lattice is $p_c=0.41$ which should equal
the density of doubly and unoccupied sites, so that $p_c=1-U/(2V)$
which implies that $(V/U)_{c}=1/{2(1-p_c)}=0.85$. 
This indicates that an estimate
of $V_{c2}\approx 3.4t$ for $U=4t$.
We have compared our results for $m^\dagger$ as a function of 
$V$ with the percolation of vacancies in a quantum Heisenberg model and find 
very good agreement \cite{chernyshev,percolation}.


Fig.~5, which shows the correlation between the strength of disorder and the 
nature of the regions, is very revealing. It clearly
shows a reorganization or a phase separation of the system
into WD regions that are AFM
and have a local density of one electron per site, and the SD regions 
which are PM with a bimodal density clustering 
towards zero and two per site.
\begin{figure}
\begin{center}
\vskip-2mm
\hspace*{1mm}
\epsfig{file=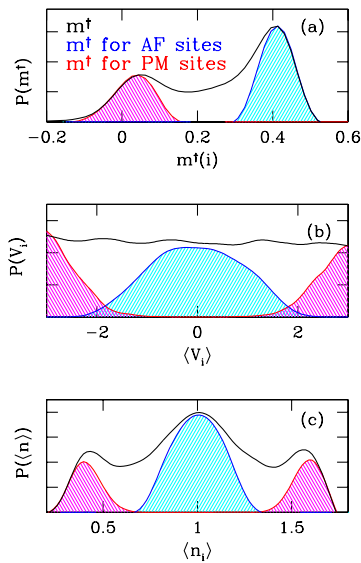,height=3.0in,angle=0}
\vskip1mm
\caption{
(a) Histogram $P(m^\dagger)$ of the local staggered magnetization
(black). We will define those sites 
with staggered magnetization 0.38 and more to be AFM sites 
(region shown in blue); and sites with $\la S_z(i) \ra$ in the range 
(-0.1,0.1) to be paramagnetic PM sites (shown in red).
(b) Histogram P(V) of the local disorder potential for all sites (black); 
histogram of AFM sites (blue) which coincide with the less disordered WD 
regions; and histogram of PM 
sites (red) in the SD regions.
(c) Histogram P(n) of the local density for all sites (black); 
for AFM sites (blue); and for PM sites (red).
}
\label{fig5}
\end{center}
\end{figure}
Our results for the existence of metallic behavior in 2D is even more 
surprising when we recall the scaling theory of localization
for non-interacting electrons\cite{gangof4}
which claims
that in dimensions less than and equal to two
all the single particle states are localized by arbitrarily small amounts of 
disorder. 
Thus the underlying picture that emerges for the origin of the unusual metal
at intermediate disorder in a Mott insulator at half filling is 
a percolation of the 
defected SD regions with reduced AFM.  
As seen in Fig.~3(a)(top panel) the lowest excitations live in these 
defected SD regions
and with increasing disorder the defective regions expand 
and along with that the excitations get more 
delocalized as they are able to travel on the expanding SD 
clusters, see Fig.~3(a) (middle panel).
This is
a rather remarkable situation that increasing disorder in a correlated 
system is able to induce metallic behavior. 
With further increase of disorder the states on the percolating cluster get
localized and the 
system becomes an Anderson insulator, see Fig.~3(a) (bottom panel).

\medskip

\ni{\it Acknowledgments:} 
NT gratefully acknowledges the 
hospitality of the Physics Department at University of Illinois and
support through DOE grant DEFG02-91ER45439 and DARPA grant N0014-01-1-1062.
DH would like to acknowledge partial support from the Kanwal Rekhi 
scholarship administered by the TIFR Alumni Association.
We acknowledge the use of computational 
facilities at TIFR.
We would also like to thank M. Barma, L.~H. Greene,
A.~J. Leggett, and M. Randeria for many useful discussions.


\begin{thebibliography}{999}
\bibitem{footnote_mu} For a given realization of the disorder potential,
the chemical potential $\mu$ is also adjusted to satisfy the half filling 
condition exactly.
\bibitem{footnote_states} The 
entire spectrum is affected by disorder.
For $V\sim t$ there is a gap
and the most 
localized states are at the edges of the gap.
As $V\sim 2t$, the gap closes and the most localized states move to the 
Fermi energy.
For $V\sim 3t$ in region II there is a dramatic change 
in IPR and the states
become extended. For higher values of $V\sim 5t$ in region III all the states,
including those at the edges of the spectrum get localized.
\end{thebibliography}
\end{document}